# Practical Evaluation of the Crypto-Agility Maturity Model

Leonie Wolf ✉[1,2,3][0009-0008-1554-9975], Samson
Umezulike[1,2,3][0009-0009-7723-1281], Gurur Öndarö[4][0009-0009-1697-5797],
Sebastian Schinzel[1,2,3,4][0000-0002-7944-5488], and Fabian
Ising[1,2,3][0000-0001-9852-1231]

[1] Fraunhofer SIT
[2] National Research Center for Applied Cybersecurity ATHENE
[3] HNFIZ Cybersecurity
[4] FH Münster University of Applied Sciences
`leonie.wolf@sit.fraunhofer.de`,
`{umezulike,gurur.ondaro,schinzel,f.ising}@fh-muenster.de`

**Abstract.** Cryptographic agility is a key prerequisite for maintaining the long-term security of digital communication, particularly in light of the transition to post-quantum cryptography. To systematically assess this capability, Hohm et al. proposed the Crypto Agility Maturity Model (CAMM). In this work, we present the first evaluation of the CAMM against established design principles for maturity models. Our analysis reveals that the CAMM only partially satisfies these principles: its scope and target groups remain ambiguous; acceptance criteria are insufficiently operationalized, limiting verifiability and replicability; and dependency relations exhibit redundancies, cycles, and omissions. Applying the CAMM to a simple real-world scenario further confirmed these issues, as several requirements at higher maturity levels proved inapplicable or unclear. Based on these findings, we propose concrete improvements to the CAMM to enable more consistent and reliable assessments of cryptographic agility.

## 1 Introduction

Cryptographic algorithms are the cornerstone of confidential communication. However, from time to time, the security of these algorithms comes under scrutiny, or more efficient algorithms are published, requiring a transition to new algorithms. Cryptographic Agility (or "crypto agility") describes the ability to switch between cryptographic algorithms and primitives. Especially as part of the transition to post-quantum cryptography, crypto agility has been endorsed by legislative bodies worldwide, such as the European Union [8] and the National Institute of Standards and Technology (NIST) [20].

However, it is challenging to quantify the degree of cryptographic agility a system or organization exhibits. To this end, in 2023, Hohm et al. proposed the Crypto-Agility Maturity Model (CAMM) [13] to systematically measure the cryptographic agility of an IT landscape. Since then, this maturity model has received public attention and has even been endorsed by the NIST [4].

This endorsement by standardization organizations, combined with the fact that it is the only peer-reviewed maturity model available for cryptographic agility, makes it reasonable to assume that in the future, governmental contracts might require a specific maturity level in the CAMM. Therefore, the CAMM should provide robust and logical evaluation criteria, and needs to undergo rigorous assessment.

In this paper, we analyze the quality and practical applicability of the CAMM. To this end, we evaluate it using criteria from previous work [24] to determine whether it fulfills the fundamental design principles of a maturity model. Moreover, we construct a real-world scenario based on a single, simple organization and try to apply the CAMM to their IT landscape.

The primary question we seek to answer in this paper is: *Is the CAMM a useful and complete maturity model for cryptographic agility and does it allow interested parties to efficiently/easily apply it to their systems?* To answer this broad question, we define the following smaller research questions:

**RQ1:** Does the CAMM fulfill the fundamental design principles for a useful maturity model?

**RQ2:** Can the CAMM be used to assess the cryptographic agility of a common scenario?

**RQ3:** What improvements can be made to the CAMM to support stakeholder in evaluating their systems?

We find that applying the CAMM to our chosen scenario is not straightforward and requires extensive discussion among researchers for some of the Requirements. We attribute this to the fact that the CAMM violates basic design principles for maturity models, and give recommendations for its improvement.

In summary, our contributions are:

– We present the first evaluation of the CAMM against fundamental design principles for maturity models as outlined by Pöppelbuß et al. [24].
– We apply the CAMM in its current state to a simplified yet realistic scenario, describing the challenges and results that arise.
– We provide recommendations to enhance the CAMM and improve its applicability.

*Limitations and future work*  The statistical significance of our practicability evaluation is limited by the number of researchers involved (i.e., 4). Moreover, our scenario is, by choice, a simplified scenario. On the one hand, choosing a more complex, real-world scenario could have resulted in more nuanced evaluation results. On the other hand, this would have significantly increased the complexity of the discussion and results. We believe that for an initial evaluation of the CAMM, our scenario is sufficient to reveal its major controversial points.

Lastly, we do not present an improved version of the CAMM in this work. From our point of view, our research could be the first step in a practical iterative improvement approach. Interesting next steps could be an improved version of the CAMM, an evaluation with more participants, and a more complex scenario. This is left to future work.

## 2  Background

### 2.1  Cryptographic Agility

The notion of cryptographic agility has been widely discussed in the literature [1][14][19]. Early references appear in RFC 6421 [18], where the term cryptographic agility was discussed in the context of RADIUS. Today, the term has become especially prominent in the field of Post-Quantum Cryptography (PQC), where the anticipated need to replace classical algorithms underpins its urgency [2][5][9][10]. The importance of crypto agility has been further emphasized by NIST, which published a white paper in July 2025 [4], that presents the technical and organizational aspects of achieving crypto agility. Similar efforts started in the European Union with the Coordinated Implementation Roadmap for Transition to PQC [8], where crypto agility is to be facilitated by 2026 and achieved by the end of 2030.

Despite this attention, there is no unified definition of crypto agility, and its meaning often depends on context [1]. In general, it can be understood as a generalization of migration, i.e., the systematic ability to adapt and replace algorithms and protocols—particularly in PQC-related research—holistic frameworks and automation tools to realize agility across complex systems remain scarce [1].

For this paper, we adopt the definition of Näther et al. [19]: "Cryptographic Agility is a theoretical or practical approach, objective, or property which provides capabilities for setting up, identifying, and modifying encryption methods and keying material in a flexible and efficient way while preserving business continuity."

### 2.2  Maturity Models

Maturity models offer structured frameworks for assessing and guiding organizational development across various domains. Originating from the Capability Maturity Model (CMM) in the 1990s [23], they typically define a sequence of levels ranging from non-existent to optimized [3]. Each level represents a characteristic set of practices and capabilities that indicate the degree of maturity of a given entity, such as processes, systems, or governance structures. While the basic purpose of a maturity model is to describe stages and maturation paths, they can further be distinguished by their specific purposes. *Descriptive* maturity models are used to determine the current maturity of the analyzed entity, producing an informative assessment of the as-is state [24]. *Prescriptive* models additionally provide the means to identify desirable maturity levels and offer specific and detailed suggestions on how to achieve them [24]. *Comparative* models support benchmarking to allow the comparison of maturity levels across similar business units or organizations [24]. Maturity models are applied in domains such as IT management [3], business process management [6], and Industry 4.0 [7]. Their adoption has recently extended to security-critical areas, including cryptographic agility [13].

### 2.3 Related Work

Existing research on maturity models spans model design, evaluation criteria, and domain-specific applications. Early work by Becker et al. [3] introduced a systematic, iterative framework for designing and evaluating maturity models, particularly within IT management, providing one of the first structured approaches in this field. Building on such generic guidelines, Otto et al. [22] proposed an eight-step process, explicitly tailored to prescriptive maturity models, with a stronger emphasis on continuous evaluation and addressing criticisms of earlier frameworks for lacking rigor and theoretical grounding. Subsequent studies [24] derived general design principles for maturity models, justified by existing literature and grouped by typical purposes of use, and serving as a checklist for designing or evaluating maturity models. Additional contributions [15,17] discussed typical phases of maturity model development. Comparative reviews [7] highlighted frequent shortcomings when scoring existing models against explicit design principles.

Within the security domain, and particularly in cryptography, the concept of cryptographic agility has gained prominence due to emerging threats such as quantum computing [2]. Research has reviewed current definitions and addressed conceptual foundations of crypto agility [19], surveyed the state of adoption and best practices [1], and proposed assessment frameworks, such as the Crypto Agility Risk Assessment Framework (CARAF) [16] and sector-focused frameworks for implementing crypto agility [9].

The Crypto-Agility Maturity Model (CAMM) [13] introduces one of the first prescriptive maturity frameworks tailored to cryptographic agility, with five levels (Initial/Not Possible, Possible, Prepared, Practiced, and Sophisticated) and reporting initial expert validation. However, unlike long-standing domains where models have undergone multiple evaluation cycles, the practical applicability and completeness of the CAMM remain underexplored. Building on design-science guidance for maturity models and comparative scoring against explicit design principles, our work evaluates the practical application of the CAMM, identifies its strengths, and pinpoints areas for improvement.

## 3 High-level Evaluation

Pöppelbuß and Röglinger were the first to propose a general framework of design principles for maturity models [24], based on an extensive review of maturity-model related literature. These design principles allow an assessment of the quality of a given maturity model, i.e., how useful it is for its intended application domain and purpose of use. We begin our evaluation of the CAMM by applying this framework. While Pöppelbuß et al. [24] clearly state that not every maturity model must meet all design principles, this analysis can reveal issues in a model's design that may reduce its practical applicability.

In addition to [13], the authors of the CAMM provide a website [11] with additional information. We use both sources of information for the analysis. The design principles proposed in [24] are grouped into three levels:

1. Basic Design Principles
2. Design Principles for Descriptive Purpose of Use
3. Design Principles for Prescriptive Purpose of Use

From [13], we assume that the CAMM is mainly intended as a descriptive model, although prescriptive properties are also occasionally suggested. However, during evaluation, it became clear that none of the Design Principles (DPs) of level three apply to the CAMM. A summary of the evaluation results against the first two levels is shown in Table 1.

### 3.1   Basic Design Principles

These are fundamental principles that any maturity model should follow to fulfill their basic purpose of describing maturity stages and paths.

**Basic information (DP 1.1)** The first Design Principle (DP) demands that a maturity model provides a set of basic information. This includes the application domain and any prerequisites for applicability, the purpose of use, the target group (i.e., the people who need to apply the model or understand its results), and the class of entities under investigation. Additionally, to facilitate comparison with related maturity models, the model should clearly differentiate itself from these models and document the design process, as well as the extent of empirical validation.

*Application Domain and Prerequisites for Applicability (1.1a), and Entities under Investigation (1.1d)* The application domain of the CAMM is described as "crypto agility in IT systems." To define "crypto agility", the authors compile a collection of desirable system properties commonly associated with the term, such as the ability to replace cryptographic algorithms with little effort and without sacrificing interoperability. A clear and formal definition of "crypto agility" is missing. However, such a definition might not be possible or sensible, given the vast amount of literature with different understandings of the term, as discussed in Section 2.1.

The model also does not provide a more specific definition or constraints for the vague application domain of "IT system." While others, such as [21] suggest scoped domains ranging from *an algorithm* over *an application* to *a complex vertical domain*, the CAMM does not define such a scope. Furthermore, there are no technical or organizational prerequisites for applying the model. Instead,

| Design Principle | Fulfillment |
|---|:---:|
| **1.1 Basic information** | |
| a) Application domain and prerequisites for applicability | ∼ |
| b) Purpose of use | ∼ |
| c) Target group | ✗ |
| d) Class of entities under investigation | ✗ |
| e) Differentiation from related maturity models | ✓ |
| f) Design process and extent of empirical validation | ✓ |
| **1.2 Definition of central constructs related to maturity and maturation** | |
| a) Maturity and dimensions of maturity | ✓ |
| b) Maturity levels and maturation paths | ✓ |
| c) Available levels of granularity of maturation | ✓ |
| d) Underpinning theoretical foundations with respect to evolution and change | ✗ |
| **1.3 Definition of central constructs related to the application domain** | ✗ |
| **1.4 Target group-oriented documentation** | ✗ |
| **2.1 Intersubjectively verifiable criteria for each maturity level and level of granularity** | ✗ |
| **2.2 Target group-oriented assessment methodology** | |
| a) Procedure model | ✗ |
| b) Advice on the assessment of criteria | ✗ |
| c) Advice on the adaptation and configuration of criteria | ∼ |
| d) Expert knowledge from previous application | ✗ |

**Table 1.** Fulfillment of design principles from [24] in the CAMM. Design principles for prescriptive maturity models are not applicable and thus excluded. ✓: Fulfilled. ✗: Not fulfilled. ∼: Partially fulfilled.

[13] suggests a uniform application of the CAMM to "IT infrastructure" and "IT landscapes", as well as to "software", "libraries", and "frameworks". For the most part, the term "system" is used to refer to the application target, suggesting unbounded applicability to any system that uses information technology.

We conclude that the application domain of the CAMM is ambiguously defined, which may hinder stakeholders in determining the model's suitability for their specific context.

*Purpose of Use (1.1b)* The purpose of use is stated as the assessment and, possibly, the development of crypto agility. Still, the descriptions are imprecise and do not clearly specify the level of assessment or guidance expected from the model.

*Target Group (1.1c)* Similarly, the target group is not explicitly defined. IT Managers are stated to be at least part of the target group. However, the in-terviews for empirical validation involved one Security Officer and one Software Architect, suggesting a larger intended target group.

*Differentiation from Related Maturity Models (1.1e) and Design Process (1.1f)* There is a clear distinction from other maturity models—the authors argue that the CAMM is the first model tailored to crypto agility. The design process follows Becker et al. [3] and is documented precisely, thus fulfilling DP 1.1f.

**Definition of Central Constructs Related to Maturity and Maturation (DP 1.2)** The second basic design principle requires that the meaning of maturity itself needs to be defined in relation to the class of entities and the application domain. This may be achieved through one or more maturation paths consisting of ordered maturity levels and their logical relationships.

*Maturity and Dimensions of Maturity (1.2a), Maturity Levels and Maturation Paths (1.2b) and Available Levels of Granularity of Maturation (1.2c)* The CAMM consists of a single maturation path that defines maturity along the dimension of "crypto agility". The maturation path is divided into five levels:

*0: Initial/Not Possible:* Crypto agility is entirely absent due to technical or structural limitations, such as hard-coded cryptography or legacy systems.
*1: Possible*: The system meets baseline design conditions that make crypto agility feasible in principle, though no active measures have been implemented.
*2: Prepared*: Initial steps toward crypto agility have been taken, with some supporting mechanisms in place, but further effort is still required for execution.
*3: Practiced*: Crypto agility is actively realized and supported through tested migration mechanisms and suitable hardware/software capabilities.
*4: Sophisticated*: Crypto agility is fully integrated, scalable, and automated across infrastructures, enabling rapid and secure cryptographic transitions.

According to [24], a high level of abstraction, such as the one provided by the CAMM through these five levels, is suitable for comparing and documenting maturity levels. The CAMM further provides a lower level of abstraction in the form of specified Requirements that comprise the levels. The Requirements are identified by a label consisting of the letter "R", followed by the number of the corresponding maturity level and a sequential ID, which does not imply any ordering or priority, separated by a dot. For instance, the Requirement "Cryptography inventory" with the label *R1.4* is the fourth Requirement of the maturity level *1: Possible*. To reach any maturity level M, all Requirements *R*M.x must be fulfilled. These Requirements are also structured into an implicit graph via dependencies; for example, the Requirements *R2.1* and *R3.8* depend on *R1.4*. The CAMM website [11] contains further properties for each Requirement besides references, category, name, and label/ID. The specific meaning of these properties is not defined, but we infer the intended meaning from their names and actual values.

- "Description": A brief high-level description of the desired state. Outliers: *R1.0*, which instead motivates the Requirement, and *R3.7*, which describes the required artifact.
- "Problem": A motivation of the Requirement. This is either phrased as a normative statement (*R1.1, R1.2, R2.3, R2.4, R3.5, R3.7, R4.0, R4.1, R4.3, R4.4*), a necessary condition for a desired property (*R1.4, R2.2, R3.3*), or describes a risk (*R2.5, R3.0, R3.1, R3.8, R4.2*) or constraint (*R1.0, 2R.1, R3.2, R3.4, R3.6*) of not fulfilling the Requirement. Outliers: *R1.3* and *R2.0*, which are phrased as desirable properties.
- "Acceptance": An abstract statement about the state of the system which must be true for the Requirement to be fulfilled. Outliers: *R3.6, R4.1*, which are phrased as normative statements.
- "Example": A more concrete example of a scenario for which the Requirement is fulfilled (*R1.0, R1.1, R1.2, R1.3, R1.4, R2.0, R2.1, R2.2, R2.4, R3.0, R3.2, R3.3, R3.5, R3.6, R3.7, R4.0, R4.1, R4.2, R4.3, R4.4*), or a negative example for which it is not (*R1.0, R1.1, R1.2*), or a more concrete instance of the problem class addressed by the Requirement (*R2.3, R2.5, R3.1*). Two outliers are *R3.4*, which suggests using the CAMM itself to fulfill the Requirement, and *R3.8*, which references a definition of the problem class.

*Underpinning Theoretical Foundations with Respect to Evolution and Change (1.2d)* The final aspect of this DP highlights the need for maturity models to explicate the underlying theoretical foundations of change in the respective field, including typical developments, drivers, and barriers of maturity [24]. The CAMM does not explicitly provide any such information.

**Definition of central constructs related to the application domain (DP 1.3)** The goal of this design principle is to enhance both "understandability" and "language adequacy." As discussed in (DP 1.1), we find the application domain of the CAMM to be ambiguously defined. Naturally, the CAMM does not include definitions of central constructs in the application domain, except those related to crypto agility.

**Target Group-oriented Documentation (DP 1.4)** Since different target groups require different levels and types of detail, (DP 1.4) requires that the documentation be composed in a way oriented towards the target group. For instance, technology-oriented audiences may need sufficient detail to enable the described artifact to be implemented. In contrast, management-oriented audiences may need sufficient detail to determine if the organizational resources should be committed to constructing or purchasing and using the artifact within their specific organizational context [12]. Since the target group itself is not clearly specified in the CAMM, the documentation cannot be target-group-oriented.

The Requirements, in general, do not include much detail of any specific kind. The authors themselves describe the Requirements as "rather abstract" and "generic" and suggest using the SMART method for specific implementation. This is also relevant for both descriptive and prescriptive purposes.

### 3.2  Descriptive Design Principles

The second set of DPs is formulated for maturity models that serve a descriptive purpose, i.e., as diagnostic tools.

**Intersubjectively Verifiable Criteria for each Maturity Level and Level of Granularity (DP 2.1)** This DP requires the model to provide assessment criteria for each maturity level and available level of granularity, specifically for the Requirements in the CAMM. These criteria should be described precisely to enable high intersubjective verifiability and ensure comparability between assessments [24].

Due to the generic definition of the Requirements in the CAMM, intersubjective verifiability is hardly achievable, as all Requirements are at least partially subjective. The "Description", "Problem", or "Example" properties provide context and additional information, rather than serving as assessment criteria. This leaves the "Acceptance" property of a Requirement as the one that is most likely to be helpful in a systematic assessment or verification thereof; however, this property is also vague or subjective across all Requirements. For instance, the "Acceptance" property of Requirement *R1.0 System Knowledge* states that it is fulfilled if "an in-depth understanding of the structure and operation of the systems being evaluated is available". However, both the degrees of "in-depth understanding" and "availability" leave substantial room for subjective interpretation, and different assessors may easily disagree on whether the Requirement is adequately fulfilled.

**Target Group-oriented Assessment Methodology (DP 2.2)** Fulfilling this DP assures that the results from an assessment are correct, accurate, and repeatable. To achieve this, the model needs to guide users through maturity assessments using a procedure model that elaborates on the assessment steps, their interplay, and, particularly, how to elicit the values of the criteria. Further, the model should provide advice on how to adapt the criteria and report from previous applications [24].

The CAMM does not offer these elements—the Requirements themselves are not intersubjectively verifiable, and there is no guidance on how to apply them. The only advice regarding adaptation is to use the SMART method to concretize Requirements.

### 3.3  Prescriptive Design Principles

The final set of Design Principles is intended for prescriptive models. While the CAMM seems to be mainly intended for descriptive purposes, the following statements suggest that it can also fulfill a prescriptive purpose (emphasis added):

- "Based on our model, the cryptographic agility of an IT landscape can be systematically measured and *improved step by step*." [13, Abstract]
- "With CAMM at hand, IT managers can systematically assess their IT infrastructure and *derive concrete measures to further develop their IT landscape* in the direction of crypto-agility." [13, Sec. 1]

However, the DPs all relate to improvement measures provided by the model. Briefly summarized, DP 3.1 requires that the provided improvement measures cover all maturity and granularity levels, DP 3.2 demands a decision calculus for selecting improvement measures, and DP 3.3 further requires the model to define a target-group-oriented decision methodology. The CAMM does not include any improvement measures, so these DPs are not applicable, indicating that the CAMM, in its current state, is not usable for a prescriptive purpose.

## 4 Evaluation of Applicability

In addition to the analysis of design principles presented above, we apply the CAMM to a simple yet practical use case to evaluate its ease of application.

### 4.1 Scenario Description

Our objective is to apply the maturity model to a practical example and map the Requirements of each maturity level onto this scenario. To keep the evaluation manageable, we deliberately limit our example to a simple setup.

We consider an HTTPS server that operates internally within an organization and provides a website to its employees. The server is physically isolated from the public internet. All HTTP connections are secured via TLS (HTTPS), and employees access the service using web browsers on their individual workstations. The employees use a standard web browser specified by the company. We assume that both the browsers and the operating systems on employee machines are regularly updated.

We consider our selected setup to be a representative scenario for assessing cryptographic agility, as it is both manageable and reflects a standard practical deployment. Since the primary goal of this section is to evaluate the CAMM's applicability, a simplified scenario is particularly advantageous.

### 4.2 Practical Application of the CAMM to the Scenario

When applying the CAMM to the scenario, our objective was not to assess the cryptographic agility of the scenario itself, but to evaluate the maturity model and its Requirements.

To this end, four researchers independently applied the CAMM to the scenario. For each Requirement, they classified whether (a) the organization could have made a decision such that the Requirement can be met, (b) the Requirement lies outside of the organization's influence, or (c) the Requirement does not apply to the scenario. The researchers also recorded any issues or ambiguities they encountered during the application of any specific Requirement. The results were then discussed, and a consensus was reached for each Requirement. An overview of this evaluation is shown in Table 2, while Table 3 shows the issues in Requirements identified through consensus.

Overall, the test scenario successfully fulfilled all Requirements of the first three levels (up to *Practiced Cryptographic Agility*). Only some Requirements at maturity level 4 (*Sophisticated*) were not applicable. Notably, when TLS was used, Requirements such as *R1.4, R2.0, R2.1, R2.2,* and *R4.2* were automatically satisfied, making it impossible not to meet them in this context.

| | **Browser** | $P_i$ | **Server** | $P_i$ | **Scenario** | $P_i$ |
|---|---|---|---|---|---|---|
| **Level 1: possible** | | | | | | |
| R1.0 System knowledge | ✓ | 1.0 | ✓ | 1.0 | ✓ | 1.0 |
| R1.1 Updateability | ✓ | 1.0 | ✓ | 1.0 | ✓ | 1.0 |
| R1.2 Extensibility | ✓ | 1.0 | ✓ | 1.0 | ✓ | 1.0 |
| R1.3 Reversibility | ✓ | 1.0 | ✓ | 1.0 | ✓ | 1.0 |
| R1.4 Cryptography inventory | ✓ | 1.0 | ✓ | 1.0 | ✓ | 1.0 |
| **Level 2: prepared** | | | | | | |
| R2.0 Cryptographic modularity | ✓ | 1.0 | ✓ | 1.0 | ✓ | 0.75 |
| R2.1 Algorithm IDs | ✓ | 1.0 | ✓ | 1.0 | ✓ | 1.0 |
| R2.2 Algorithm intersection | – | 0.0 | – | 0.0 | ✓ | 1.0 |
| R2.3 Algorithm exclusion | ✓ | 0.5 | ✓ | 1.0 | ✓ | 1.0 |
| R2.4 Opportunistic security | ✓ | 1.0 | ✓ | 1.0 | ✓ | 1.0 |
| R2.5 Usability | ✓ | 0.5 | ✓ | 0.5 | ✓ | 0.5 |
| **Level 3: practiced** | | | | | | |
| R3.0 Policies | ✓ | 0.75 | ✓ | 1.0 | ✓ | 1.0 |
| R3.1 Performance Awareness | ✓ | 0.25 | ✓ | 0.5 | ✓ | 0.5 |
| R3.2 Hardware Modularity | ✓ | 0.75 | ✓ | 0.75 | ✓ | 0.75 |
| R3.3 Testing | – | 0.25 | – | 0.25 | ✓ | 0.75 |
| R3.4 Enforceability | O | 0.5 | ✓ | 0.5 | ✓ | 0.75 |
| R3.5 Security | ✓ | 0.0 | ✓ | 0.0 | ✓ | 0.25 |
| R3.6 Backwards Compatibility | ✓ | 0.75 | ✓ | 0.75 | ✓ | 0.75 |
| R3.7 Transition Mechanism | ✓ | 1.0 | ✓ | 1.0 | ✓ | 1.0 |
| R3.8 Effectiveness | – | 0.25 | ✓ | 0.75 | ✓ | 0.75 |
| **Level 4: sophisticated** | | | | | | |
| R4.0 Automation | ✓ | 1.0 | ✓ | 1.0 | ✓ | 1.0 |
| R4.1 Context Independence | – | 0.75 | – | 0.75 | – | 1.0 |
| R4.2 Scalability | – | 0.75 | – | 0.75 | – | 1.0 |
| R4.3 Real-Time | ✓ | 0.75 | ✓ | 1.0 | ✓ | 1.0 |
| R4.4 Cross-System Interoperability | ✓ | 0.0 | ✓ | 0.0 | ✓ | 0.0 |

✓  The organization could have made decisions such that it is possible to meet this Requirement.
O  It is out of the organization's scope of responsibility to fulfill this Requirement.
–  The Requirement is not applicable to the scenario.
The $P_i$ values represent the proportion of individual ratings agreeing with the final consensus for each item. They are illustrative only and are presented to indicate the relative ease of achieving consensus.

**Table 2.** Application of the CAMM to the test scenario

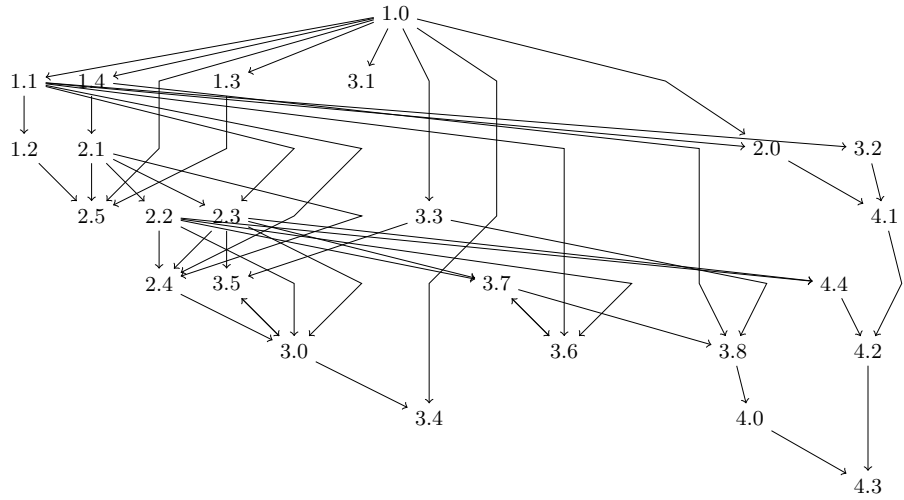

**Fig. 1.** Dependencies as depicted in [13]
.

### 4.3   Observations

Our evaluation revealed several recurring issues with the CAMM Requirements, which we categorize into four areas: recursive references, measurability problems, unclear or undesirable Requirements, and dependency inconsistencies. A detailed overview is provided in Table 3.

**Recursive Requirements** Several Requirements (e.g., *R3.1*) refer back to the term "crypto agility" itself. Since the CAMM is designed to encompass the various aspects of crypto agility and provide a basic understanding of it, using the keyword it describes leads to recursion and reduces clarity.

**Measurability Problems** As described in the previous section, maturity models should include verifiable criteria for each requirement. However, several requirements are vague and lack measurable acceptance criteria.

For instance, Requirement *R1.0 System Knowledge* leaves open the question of how much knowledge is "enough" to meet the requirement. Similarly, *R2.5 Usability* should, according to [11], be accepted when a usability study demonstrates that the system is easy to use. Yet the model does not specify for whom it must be easy to use, nor how such a study should be designed. These gaps make consistent evaluation difficult. Requirement *R3.5 Security*: As shown in Table 2, the $P_i$s for this requirement are fairly low, i.e., there was little agreement on our initial ratings compared to the final consensus. The discussion to reach a final consensus focused on different interpretations of the text, especially of the acceptance criterion. In this case, it neither defines the kind nor the level of security needed to fulfill this Requirement. This leaves a lot of room for interpretation. Furthermore, since security is a general requirement for encrypted communication, it is not a measure of cryptographic agility. Even if included, this requirement should not be considered as late as level 3 of the model. Lastly, *R4.3 Real-Time* leaves the specification of "Real-Time" in this context up for interpretation. Since it is not defined, this Requirement is also difficult to distinguish from *R3.8 Effectiveness*.

**Unclear and Undesirable Requirements** Since the CAMM is intended as a general-purpose model for assessing cryptographic agility, all of its requirements must be applicable and desirable across all scenarios within its scope. During the practical application of the CAMM to our test scenario, we identified requirements that, in our opinion, should not be included as necessary criteria for cryptographic agility.

*Fundamental Properties rather than Agility Criteria* Some requirements describe basic features of any communication system rather than crypto agility specifically. One example of this is Requirement *R2.2 Algorithm Intersection*, a basic foundation for encrypted communication. While it is essential to ensure that this fundamental property is not lost due to crypto agility, this requirement is already covered by Requirements *R3.6 Backwards Compatibility* and *R3.7 Transition Mechanism*. These two requirements, as specified in [13], on the other hand, do not refer to cryptography at all but specify general compatibility requirements for IT systems during updates.

In contrast to many other Requirements, *R3.1 Performance Awareness* is a business requirement rather than a cryptographic or technical requirement. While it is essential to understand the performance impact of implementing crypto agility, it is irrelevant to systems that already employ some agility mechanism.

*Undesirable Requirements* For some Requirements, we suggest either removing them from the model completely or at least changing them. One example is *R2.4 Opportunistic Security*, which may lead to unencrypted communication or downgrade to less secure algorithms. Depending on the use case and security considerations, there may be good reasons for opportunistic security, such as improved connectivity, but also reasons against it (e.g., confidentiality concerns). It should not be mandatory to implement opportunistic security to be crypto agile.

At level 4, the Requirements *R4.1 Context Independence*, *R4.2 Scalability*, and *R4.4 Cross-System Interoperability* have no impact on the system's crypto agility. A possible effect on the "global IT infrastructure" is—depending on the target group—not desirable. For instance, in environments with varying constraints, these Requirements cannot be fulfilled.

For *R4.0 Automation*, we question whether decisions about cryptography should be made without human interaction, and how these decisions should be made. Therefore, we believe that it should not be a requirement for crypto agility.

**Dependency Inconsistencies** The CAMM contains three types of dependencies between Requirements. First, all Requirements explicitly state dependencies in the "Dependency" property. Second, since these Requirements may depend on different ones, the explicit links introduce additional implicit transitive dependencies. Finally, because a maturity level can only be reached if all Requirements of that level are fulfilled, each Requirement implicitly depends on all Requirements at lower levels.

Constructing the dependency graph from the explicit dependencies revealed several inconsistencies that complicate its interpretation. These findings, along with potential resolutions, are outlined below.

*Single Root* The dependency graph has a single root, Requirement *R1.0 System Knowledge*. Many Requirements depend directly and solely on it, and since the graph is connected, every single Requirement ultimately depends on *R1.0*. Separating this Requirement into a Level 0 to highlight its foundational role and importance might improve the graph's structure.

*Redundant and Implicit "Short-Cuts"* Some Requirements include redundant explicit dependencies. For example, *R2.5* depends on *R2.1*, which in turn depends on *R1.4*, which depends on *R1.0*. It follows directly that *R2.5* can only be fulfilled if *R1.0* is fulfilled. Regardless, the dependency of *R2.5* on *R1.0* is stated explicitly, while the other implicit dependency on *R1.4* is omitted. These inconsistencies complicate the dependency graph. To improve clarity, future versions should establish and consistently apply clear criteria for when dependencies are listed explicitly, and specify how these differ from implicit dependencies.

*Cyclic Dependencies* The graph is mostly directed, with two exceptions: Requirements *R3.0 Policies* and *R3.5 Security* depend on each other, as well as *R3.6 Backwards Compatibility* and *R3.7 Transition Mechanism*. These interdependencies imply that the Requirements can only be fulfilled simultaneously, which complicates their interpretation. If they are indeed this tightly coupled, it might be better to merge each pair into a single Requirement.

*Missing Dependencies* We also found missing logical dependencies.

- Requirement *R1.3 Reversability* should depend on *R1.1 Updateability*, since reverting an update implies that updates can be applied in the first place.
- Requirement *R2.4 Opportunistic Security* should depend on *R1.4 Cryptographic Inventory*, since selecting the strongest available algorithm requires knowledge of which cryptographic functions are in use and their current security level.

| | Dependencies | Measurability | Recursion | Fundamental |
|---|---|---|---|---|
| **Level 1: possible** | | | | |
| R1.0 System knowledge | – | ✗ | – | – |
| R1.1 Updateability | – | – | – | – |
| R1.2 Extensibility | – | – | – | – |
| R1.3 Reversibility | ✗ | – | – | – |
| R1.4 Cryptography inventory | – | – | – | – |
| **Level 2: prepared** | | | | |
| R2.0 Cryptographic modularity | ✗ | ✗ | – | – |
| R2.1 Algorithm IDs | – | – | – | – |
| R2.2 Algorithm intersection | – | – | – | ✗ |
| R2.3 Algorithm exclusion | – | – | – | – |
| R2.4 Opportunistic security | ✗ | – | – | ✗ |
| R2.5 Usability | ✗ | ✗ | ✗ | ✗ |
| **Level 3: practiced** | | | | |
| R3.0 Policies | – | – | – | – |
| R3.1 Performance Awareness | – | ✗ | ✗ | ✗ |
| R3.2 Hardware Modularity | – | – | – | – |
| R3.3 Testing | – | – | – | – |
| R3.4 Enforceability | – | – | ✗ | – |
| R3.5 Security | ✗ | ✗ | – | ✗ |
| R3.6 Backwards Compatibility | ✗ | – | – | ✗ |
| R3.7 Transition Mechanism | ✗ | – | – | ✗ |
| R3.8 Effectiveness | ✗ | – | – | – |
| **Level 4: sophisticated** | | | | |
| R4.0 Automation | – | ✗ | – | ✗ |
| R4.1 Context Independence | – | ✗ | – | ✗ |
| R4.2 Scalability | – | ✗ | – | ✗ |
| R4.3 Real-Time | – | ✗ | – | ✗ |
| R4.4 Cross-System Interoperability | ✗ | – | – | ✗ |

**Table 3.** Issues identified in the CAMM Requirements (by consensus between researchers)

– Requirement *R4.4 Cross-System Interoperability* should depend on *R2.2 Algorithm intersection*, since interoperability between systems is only possible when they support a common set of cryptographic algorithms.

## 5   Recommendations for Improving the CAMM

During our high-level analysis and application of the CAMM to an example scenario, we identified several shortcomings in its design, related to both fundamental design principles and practical usage problems.

### 5.1   Clarify definitions of Scope and Application Domain

The most glaring issue of the CAMM is the omission of a clear definition of the model's scope, including both the application domain ("IT system") and the term "crypto agility", as well as an explicit definition of the target groups.

This is not only an academic issue; it also prevents potential stakeholders from applying the model at all.

For example, our research group extensively discussed how to select the scenario for a meaningful evaluation of the CAMM. While our first attempt was modeling a protocol (i.e., TLS), we quickly decided against it because many Requirements did not apply. A more detailed scenario, on the other hand, led to many Requirements that were not easily measurable. We believe the ambiguous scope and definitions of the CAMM make it challenging to apply it to anything but a trivial scenario such as the one described in Section 4.1. Therefore, one (or more) focused and narrowly defined scopes, along with an explicit definition of target groups, would improve the applicability and comparability of the CAMM.

### 5.2   Improve Measurability of Requirements

The aspect that hindered the evaluation of our scenario the most was that the Requirements were not easily measurable. Even though each Requirement comes with an acceptance criterion [11], many of these criteria are vague and hard to measure effectively. In our practical evaluation, this consistently led to discrepancies in the assessment of Requirements.

Therefore, we recommend restructuring the acceptance criteria to be more transparent and easier to measure, ideally by providing concrete, testable indicators and examples of valid assessment methods (e.g., by linking to checklists, defining metrics, or test procedures).

### 5.3   Simplify and Correct Dependencies

Transparent dependencies between Requirements are essential to allow stakeholders to define logical progression paths, i.e., which step to take next to improve an organization's cryptographic agility.

Concrete steps towards improving this are

1. Moving Requirement *R1.0 System Knowledge* to be a pre-requisite. This could be achieved by defining a Maturity Level 0.
2. Removing redundancies
3. Merging interdependent Requirements
4. Adding missing dependencies (e.g., *R1.3* should depend on *R1.1*)

### 5.4   Documentation and Guidance

To support stakeholders in applying the CAMM, we recommend adding target-group-oriented documentation, such as technical details for implementers (e.g., the specific steps to achieve acceptance of a Requirement) or high-level guidance for IT managers (e.g., the consequences of not meeting a Requirement). This is especially relevant if the CAMM is intended to be used as a prescriptive model. In this case, it should additionally provide improvement pathways, such as prioritization of Requirements and suggested next steps for each maturity level.

## 6   Conclusion

In this work, we evaluated the Crypto-Agility Maturity Model (CAMM) [13], which, among others, was mentioned by NIST as a potential tool for measuring cryptographic agility. In particular, we evaluated the model against a framework of design principles [24] and applied it to a simple, but real-world scenario.

We find that the CAMM does not fulfill many of the fundamental design principles outlined in related work (RQ1). This is reflected in the observations we made during the application of the model. While we could reach consensus among four security researchers on the fulfillment of Requirements in our scenario, in many cases, we had to discuss the acceptance criteria extensively beforehand (RQ2). We believe that standard organizations will struggle even more with evaluating their vastly more complex IT landscape against the need for cryptographic agility.

Despite these criticisms, we believe that the CAMM is a crucial step towards measuring and enhancing cryptographic agility in organizations. However, defining a general-purpose maturity model for a property as complex as cryptographic agility is non-trivial. To support this process, we recommend concrete improvements to the model (RQ3). We believe these improvements will allow more focused and effective use of the CAMM and pave the way towards the important goal of cryptographic agility.

*Acknowledgements* Gurur Öndarö was supported by the research project "North-Rhine Westphalian Experts in Research on Digitalization (NERD II)", sponsored by the state of North Rhine-Westphalia – NERD II 005-2201-0014. This research work was supported by the National Research Center for Applied Cybersecurity ATHENE and by the Dieter Schwarz Foundation.